\documentclass[prd,aps,superscriptaddress,showpacs,amssymb,
amsmath,twocolumn,floatfix]{revtex4}
\usepackage{bm,graphicx,dcolumn,amsmath,hyperref}

\newlength{\myfigwidth}
\graphicspath{{Figures/}}

\usepackage{ifpdf}
\ifpdf
\setkeys{Gin}{width=\myfigwidth,clip,angle=90}
\else
\setkeys{Gin}{width=\myfigwidth,clip}
\fi

\newcommand{\bE}{\mathbf{E}}
\newcommand{\br}{\mathbf{r}}
\newcommand{\bk}{\mathbf{k}}
\newcommand{\bzero}{\mathbf{0}}
\newcommand{\bit}{\begin{itemize}}
\newcommand{\eit}{\end{itemize}}
\newcommand{\ben}{\begin{equation}}
\newcommand{\een}{\end{equation}}
\newcommand{\bea}{\begin{eqnarray}}
\newcommand{\eea}{\end{eqnarray}}
\newcommand{\bob}{\underline{b}}
\newcommand{\bb}{\left(}
\newcommand{\eb}{\right)}
\newcommand{\la}{\left\langle}
\newcommand{\ra}{\right\rangle}

\newcommand{\rone}{\hat{r}_1}
\newcommand{\rzero}{\hat{r}_0}

\begin{document}

\title{Radiative corrections to the lattice gluon action for highly
  improved staggered quarks (HISQ) and the effect of such corrections on the
  static potential}

\author{A. \surname{Hart}}
\affiliation{SUPA, School of Physics and Astronomy, University of Edinburgh, 
Edinburgh EH9 3JZ, United Kingdom}

\author{G.M \surname{von Hippel}}
\affiliation{Deutsches Elektronen-Synchrotron DESY, 15738 Zeuthen, Germany}

\author{R.R. \surname{Horgan}}
\affiliation{Department of Applied Mathematics and Theoretical Physics,
University of Cambridge, Centre for Mathematical Sciences, Cambridge
CB3 0WA, United Kingdom}

\collaboration{HPQCD Collaboration}
\noaffiliation

\preprint{DAMTP-2008-110}
\preprint{DESY 08-181}
\preprint{Edinburgh 2008/46}

\begin{abstract}
We perform a perturbative calculation of the influence of dynamical
HISQ fermions on the perturbative improvement of the gluonic action in
the same way as we have previously done for asqtad fermions. We find
the fermionic contributions to the radiative corrections in the
L\"uscher-Weisz gauge action to be somewhat larger for HISQ fermions
than for asqtad. Using one-loop perturbation theory as a test, we
estimate that omission of the fermion-induced radiative corrections in
dynamical asqtad simulations will give a measurable effect. The
one-loop result gives a systematic shift of about $-0.6\%$ in $\rone$
on the coarsest asqtad improved staggered ensembles. This is the
correct sign and magnitude to explain the scaling violations seen in
$\Phi_B = f_B \sqrt{M_B}$ on dynamical lattice ensembles.
\end{abstract}

\maketitle

\section{Introduction}
\label{sec_intro}

The Fermilab, MILC, HPQCD and UKQCD Collaborations are involved in an
ambitious programme of high precision predictions of
phenomenologically relevant parameters from QCD using unquenched
lattice simulations~\cite{Davies:2003ik}.

Central to this programme is the perturbative improvement of the
fermionic and gluonic action and operators to remove significant
sources of scaling violation in the lattice simulation results. 
This body of work is based on the Symanzik-improved staggered-quark
formalism, specifically the use of the asqtad action
\cite{Orginos:1999cr}.
More recently, the Highly Improved Staggered Quark (HISQ) action has
been used to further suppress taste-changing interactions and to allow
the use of heavier quarks at the same lattice spacing by removing
tree-level $\mathcal{O}((ma)^4)$ artifacts from the valence quark
action
\cite{Follana:2006rc}.

To maintain the same level of improvement when these actions are used
to describe the sea quarks
\cite{Wong:2007uz,Bazavov:2009jc},
we should include the effect of fermion loops on the radiative terms
in the Symanzik--improved gauge action. This has recently been done
to $\mathcal{O}(N_f \alpha_s a^2)$ for the asqtad action
\cite{Hao:2007iz}
and in this paper we update that calculation to include dynamical HISQ
fermions instead. Some preliminary results can be found in
Ref.~\cite{Hart:2008zi}. We note that the corrections are larger for
HISQ than for asqtad.

In the second part of this paper, Sec.~\ref{sec_milc_pot}, we consider
what effect the $\mathcal{O}(N_f \alpha_s a^2)$ will have in a
practical simulation, particularly on the scale-setting parameters
$\rone$ and $\rzero$ derived from the static quark potential.

The MILC and UKQCD Collaborations have already used dynamical asqtad
quarks to generate a large set of Monte Carlo lattice ensembles,
including ones with very light sea quarks (MILC,
e.g.~\cite{Davies:2003ik}) and ones with large numbers of independent
configurations (UKQCD~\cite{Gregory:2008mn}).

The gauge action used, however, omitted the asqtad $\mathcal{O}(N_f
\alpha_s a^2)$ radiative improvements (they were not then known). It
has been observed that the quantity $\Phi_B = f_B \sqrt{M_B}$ shows a
$+2\%$ scaling violation on the dynamical ``coarse'' ensembles
(lattice spacing $a \simeq 0.12~\text{fm}$) relative to the ``fine''
($a \simeq 0.09~\text{fm}$)
\cite{Gray:2005ad}. 
This scaling violation was not seen for corresponding quenched
lattices
\cite{Davies:comm}.

Unless there is a subtle (and therefore unlikely) cancellation, the
quenched result suggests that the (quenched) gluonic and valence
staggered actions are not the problem. If the asqtad action is
suitable for the valence quarks, it seems likely it is equally
suitable for the sea quarks. The scaling violation is therefore argued
to arise from the omission of the $\mathcal{O}(N_f \alpha_s a^2)$
radiative corrections to the gluonic action. It is certainly plausible
that the fermionic contributions could have such an effect; they are,
after all, large enough to reverse the sign of some of the radiative
couplings in the action
\cite{Hao:2007iz}.

In these calculations $\rone \equiv r_1/a$ has been used to set the
scale, i.e. to convert from dimensionless lattice results to physical
predictions. We therefore attempt to estimate, at least
semi-quantitatively, whether the omission of the asqtad
$\mathcal{O}(N_f \alpha_s a^2)$ radiative corrections would have a
measurable effect on the static potential and, in particular, on the
scale-setting parameters~$\rone$ and~$\rzero$ used to convert from
dimensionless lattice results to physical predictions. We do this
using one-loop perturbation theory. We treat such a result as
indicative: we do not rule out higher loop and non-perturbative
contributions, but argue that if one-loop perturbation theory predicts
a measurable result then it is likely to persist when we include other
contributions.

Using one-loop perturbation theory, we find that including the
fermionic radiative corrections to the gauge action would lead to a
$0.65\%$ \textit{decrease} in $\rone$ on the coarse ensembles and no
change on the fine. The sign and magnitude of these shifts are robust
under reasonable variations in fitting parameters. This would equate
to a $0.65\%$ \textit{increase} in $a$ on the coarse ensembles. The
quantity $\Phi_B$ scales as $a^{-3/2}$, so the $\mathcal{O}(N_f
\alpha_s a^2)$ corrections would lead to a $1\%$ \textit{decrease} in
$\Phi_B$ on the coarse ensembles and no effect on the fine.

This shift is very close to what has been observed and we therefore
suggest that the anomalous upward shift in $\Phi_B$ is in large part
due to the omission of the $\mathcal{O}(N_f \alpha_s a^2)$ radiative
corrections to the gluon action.
We therefore predict that other observables scaling with a similar
negative power of $a$ should exhibit similar scaling violations that
ought to become noticeable if these observables are measured to similar
accuracy.


\section{On-shell improvement}

We begin by briefly reviewing how radiative improvement works at
$\mathcal{O}(\alpha_s a^2)$.

Starting from the Symanzik tree-level--improved gauge action, the
Coulomb self-energy from the 1-loop radiative corrections is (see
Section~4 of Ref.~\cite{WeiWoh:1} and Eq.~(44) of
Ref.~\cite{Snippe:1997ru}):
\ben
w(\bk)~\propto~\bk^2 + \alpha_s\bb a_1\bk^2-\beta_0\bk^2\ln(\bk^2) + 
a_2\bk^{(4)}+a_3(\bk^2)^2\eb
\label{eqn_wk}
\een
where
\ben
\bk^2 = \sum_{i=1}^3 k_i^2 \; , \quad 
\bk^{(4)} = \sum_{i=1}^3 k_i^4 \; .
\een
The first two terms in the brackets are absorbed into the scheme
definition of $\alpha_V$ (Eq.~(46) of Ref.~\cite{Snippe:1997ru}) and
do not concern us. The last two terms in the brackets are lattice
artifacts and are $\mathcal{O}(\alpha_s a^2)$. It is the goal of
radiative improvement to remove these terms.

To improve the gauge theory at $\mathcal{O}(\alpha_s a^2)$ we
introduce appropriate radiative counterterms into the gauge
action. There are four such dimension-6 counterterms: three gluonic
operators (named by L\"uscher and Weisz
\cite{Luscher:1985zq}
as ``planar rectangles'', ``parallelograms'' and ``bent rectangles'',
with coefficients $c_1$, $c_2$ and $c_3$ respectively) plus the static
quark operator,
\begin{equation}
c_4\;a^2\Psi^\dagger \nabla\cdot \bE \Psi \; ,
\label{eqn_contact}
\end{equation}
which contributes specifically to the static-quark
potential.

The action normalisation condition 
\begin{equation}
c_0 + 8(c_1 + c_2) + 16 c_3 = 1
\end{equation}
ensures we get the correct gauge action in the continuum limit and
fixes $c_0$ (the coefficient of the plaquette), given the other
coefficients.

On-shell observables will remain unchanged under field redefinitions
using the equations of motion. If we confine our attention to on-shell
quantities, we can exploit this to set one of
the $c_i$ to zero~%
\footnote{Of course, this introduces higher dimensional operators into
  the theory, both from the field redefinition and from the Jacobian,
  but such operators are irrelevant from the point of view of the
  renormalisation group
  \cite{Lepage:1996jw,Alford:1995hw}.
  Alternatively, by introducing such operators in the original theory,
  we can arrange for their coefficients to vanish after the
  redefinition.}.
The usual choice is to set $c_3=0$. 

Looking at the terms in Eqn.~(\ref{eqn_wk}) in more detail, the $a_2$
term breaks rotational symmetry and its effect on the static-quark
potential is given by the Fourier transform
\ben
\delta V_2(r) \sim \alpha_s a^2\int_{-\pi}^\pi \frac{d^3k}{(2\pi)^3} 
    e^{-\bk\cdot\br}\;\frac{\bk^{(4)}}{(\bk^2)^2} \; .
\een
The leading $\mathcal{O}(\alpha_s a^2)$ behaviour is $\sim \alpha_s
a^2/r^3$ with $\mathcal{O}(a^4)$ corrections that break rotational
symmetry.

The effect of radiative improvement on the static potential is to set
$a_2 = 0$, which therefore also restores rotational invariance of the
static potential (at this level) and gives the correct Coulomb
coefficient $\alpha_V$
\cite{Snippe:1997ru}.

The $a_3$ term, in contrast, already preserves rotational invariance:
\ben
\delta V_3(r) \sim \alpha_s a^2\int_{-\pi}^\pi \frac{d^3k}{(2\pi)^3}    
 e^{-\bk\cdot\br}~+~\mathcal{O}(a^4).
\een
with the leading contribution to the static-quark potential being the
3D Kronecker $\delta_{\br,\bzero}$ (seen by changing variables to $z_i
= e^{-k_ir_i}$). This, as Snippe points out
\cite{Snippe:1997ru}, 
does not affect the potential at non-zero $r \equiv | \br |$ and will
therefore not contribute to the scale setting parameters $\rone$
and $\rzero$. In general, however, we do need to remove it: as well as
the contact term there will be an effect for $r > 0$ at higher order
(i.e. at $\mathcal{O}(a^4)$) because the denominator in the Symanzik
tree-level Coulomb propagator will not exactly cancel the $\bk^2$ from
the Feynman rules owing to differences in their definitions.

Both $c_3$ and $c_4$ contribute to the $a_3$ term
\cite{Wei:1,WeiWoh:1}
so, with $c_3=0$ fixed as above, we can only remove it by introducing
the static quark counterterm into the theory
\cite{Lepage:1996jw,Alford:1995hw},
i.e. by choosing an appropriate, and non-zero, value for $c_4$.  This
has the effect of introducing staples onto temporal Wilson lines,
which must be included in numerical simulations. Similarly, the $c_4$
contact term will be important in, for instance, the $\Upsilon(2S-1S)$
mass splitting. A contact term gives a contribution proportional to
the square of the wavefunction at the origin. This is clearly
different for the two states concerned, and will change the mass
splitting 
\footnote{We note in passing here that an alternative to the
above procedure, we could instead choose to remove the static quark
counterterm from the outset by using the equations of motion to set
$c_4 = 0$. We would then, however, have to include a non-zero $c_3$ to
counter the effect of $a_3$ leading to the inclusion of all three
improvement terms in the gauge action.}.
We will not, however, consider the contact term in detail in this
paper.

\subsection{The calculation}

Contact term aside, with $c_3=0$ we thus need to determine $c_1$ and
$c_2$ to complete the on-shell improvement.
Given two independent quantities $Q_1$ and $Q_2$ with expansions
\begin{equation}
Q_i = \bar{Q}_i + w_i (\mu a)^2 + d_{ij} c_j (\mu a)^2 +
\mathcal{O}\left((\mu a)^4\right)\;,
\end{equation}
in powers of $(\mu a)$, where $\mu$ is some energy scale, we obtain
the $\mathcal{O}(a^2)$ matching condition
\begin{equation}
d_{ij} c_j = -w_i\;.
\label{eqn:impcond_generic}
\end{equation}
Since this equation is linear, both sides can be decomposed
into a gluonic and a fermionic part; the gluonic part is known
\cite{Luscher:1985wf,Snippe:1997ru}
and is independent of the fermion action.

In this paper, we focus on the fermionic contribution to the radiative
improvement of the gluon action. Such contributions come from quark
loops, which therefore cannot change the tree-level coefficients
compared to the quenched case
\cite{Luscher:1985wf}.
To compute the one-loop HISQ fermionic corrections to the gluon action,
we will follow the same procedure as in the case of the asqtad action
\cite{Hao:2007iz},
using as our two quantities $Q_i$ the three-gluon
coupling and the mass of the so-called twisted A meson
\cite{Luscher:1985zq}.
\begin{figure}
\begin{center}
\includegraphics[height=6.4cm,keepaspectratio=,clip=]{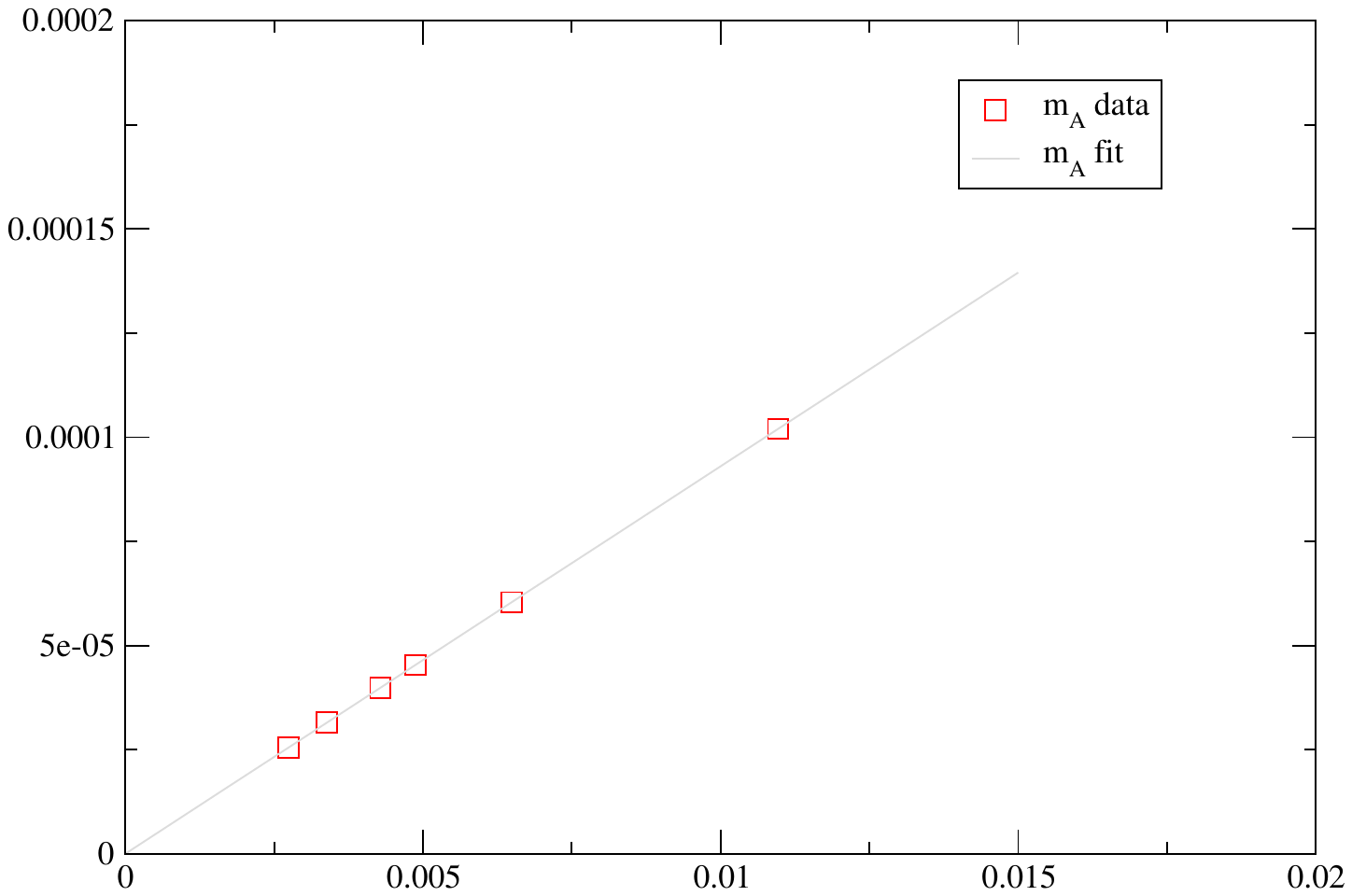}
\end{center}
\caption{\label{fig_1plot}
A plot of the fermionic contributions to the one-loop $A$
  meson self-energy $m_A^{(1)}/m$ against $(ma)^2$. The vanishing of
  $m_A^{(1)}/m$ in the infinite-volume limit can be seen clearly.}
\end{figure}

\subsection{Lattice perturbation theory}
\label{subsec:lpt}

We use lattice perturbation theory to calculate the radiative
corrections.  The (unsmeared) link variables $U_\mu$ are expressed in
terms of the gauge field $A_\mu$ as
\begin{equation}
U_\mu(x) =
\exp\left(g a A_\mu\left(x+\textstyle\frac{1}{2}\hat{\mu}\right)\right)
\end{equation}
which, when expanded in powers of $g$, leads to a perturbative
expansion of the lattice action, from which the perturbative vertex
functions can be derived.

The gauge field $A_\mu$ is Lie algebra-valued, and can be decomposed
as
\begin{equation}
A_\mu(x) = \sum_a A_\mu^a(x) t^a\;,
\end{equation}
with the $t^a$ being anti-Hermitian generators of SU($N$), where $N=3$
in the case of QCD.

The improved L\"{u}scher--Weisz action that we study
is~\cite{Luscher:1984xn}
\begin{equation}
S = 
\sum_x \left\{ c_0 P_0(x) + c_1 P_1(x) + c_2 P_2(x) \right\} \; .
\label{eqn:lw_action}
\end{equation}
with $c_0 + 8(c_1 + c_2) = 1$ and, at tree level, $c_0^{(0)} =
\frac{5}{3}$, $c_1^{(0)} = -\frac{1}{12}$, $c_2^{(0)} = 0$. The terms
\begin{align}
P_0 & = \sum_{\mu < \nu} 
  U_{\mu \nu} \; ,
\nonumber\\
P_1 & = \sum_{\mu < \nu}  \left(
  U_{\mu \mu \nu} + U_{\mu \nu \nu}  \right) \; ,
\nonumber\\
P_2 & =  \sum_{\mu < \nu < \sigma} \left(
  U_{\mu \nu \sigma} + U_{\mu \sigma \nu} + U_{\sigma \mu \nu} +
U_{\sigma; -\mu; \nu}  \right)\; ,
\end{align}
are made up of appropriate traced, closed contours of gauge links. The
notation here is that $\mu$, $\nu$ and $\sigma$ are summed over
positive values and negative subscripts denote hermitian-conjugated
gauge links.

The HISQ fermionic action is defined by an iterated smearing procedure with
reunitarisation:
\begin{equation}
U^\textrm{HISQ} = 
( F_{\textrm{asq}'} \circ P_{U(3)} \circ F_{\textrm{Fat7}} )[U]
\end{equation}
where $P_{U(3)}$ denotes the polar projection onto $U(3)$ (as used in
simulations
\cite{Bazavov:2009jc}, 
and \textit{not} $SU(3)$), and the Fat7 and modified asq smearings are
defined in
Ref.~\cite{Follana:2006rc}.

To handle the complicated form of the vertices and propagators in
lattice perturbation theory, we employ a number of automation methods
\cite{Drummond:2002kp,Hart:2004bd,Hart:prog,Nobes:2001tf,Nobes:2003nc,
Trottier:2003bw}
that are based on the seminal work of L\"uscher and Weisz
\cite{Luscher:1985wf}
and are implemented in the \textsc{HiPPy} package
\cite{Hart:2004bd,Hart:prog}.

The multi-level smearing of the gauge fields employed in the HISQ
action presents particular problems when deriving the Feynman
rules, even when employing automated techniques.  The solution to
these is discussed in Refs.~\cite{Hart:2008zi,Hart:prog}.

Unless otherwise stated, we shall use $g^2$ as the perturbative
expansion parameter (rather than $\alpha_s = \frac{g^2}{4\pi}$), with
expansions written in the form:
\begin{equation}
c_i = c_i^{(0)} + g^2 c_i^{(1)} + \mathcal{O}(g^4) \; .
\end{equation}
The goal of this paper is to determine the fermionic contributions to
$c_1^{(1)}$ and $c_2^{(1)}$, with
$c_0^{(1)} = -8 (c_1^{(1)} + c_2^{(1)})$.

Since we will only consider fermionic loops, we do not need to concern
ourselves with the gauge fixing, Haar measure and Fadeev-Popov ghost
terms that appear in the gluonic portion of the perturbative Lagrangian.

The loop integrals of continuum perturbation theory are replaced by
finite sums over the points of the reciprocal lattice in lattice
perturbation theory. We carry out these sums exactly rather than
using a stochastic estimator.


%
%
%
%
\begin{figure*}
\includegraphics[height=6cm,keepaspectratio=,clip=]{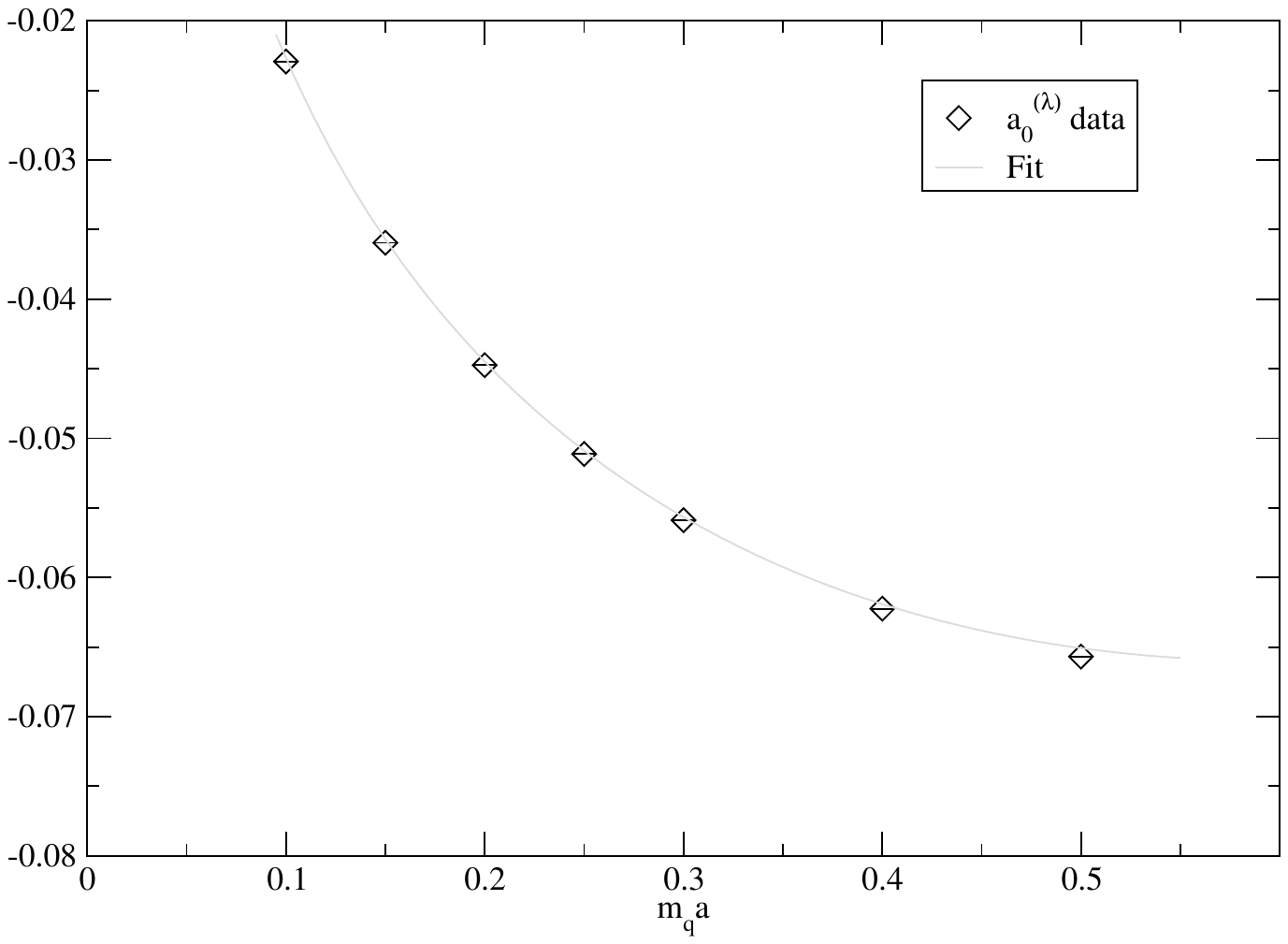}
\hfill
\includegraphics[height=6cm,keepaspectratio=,clip=]{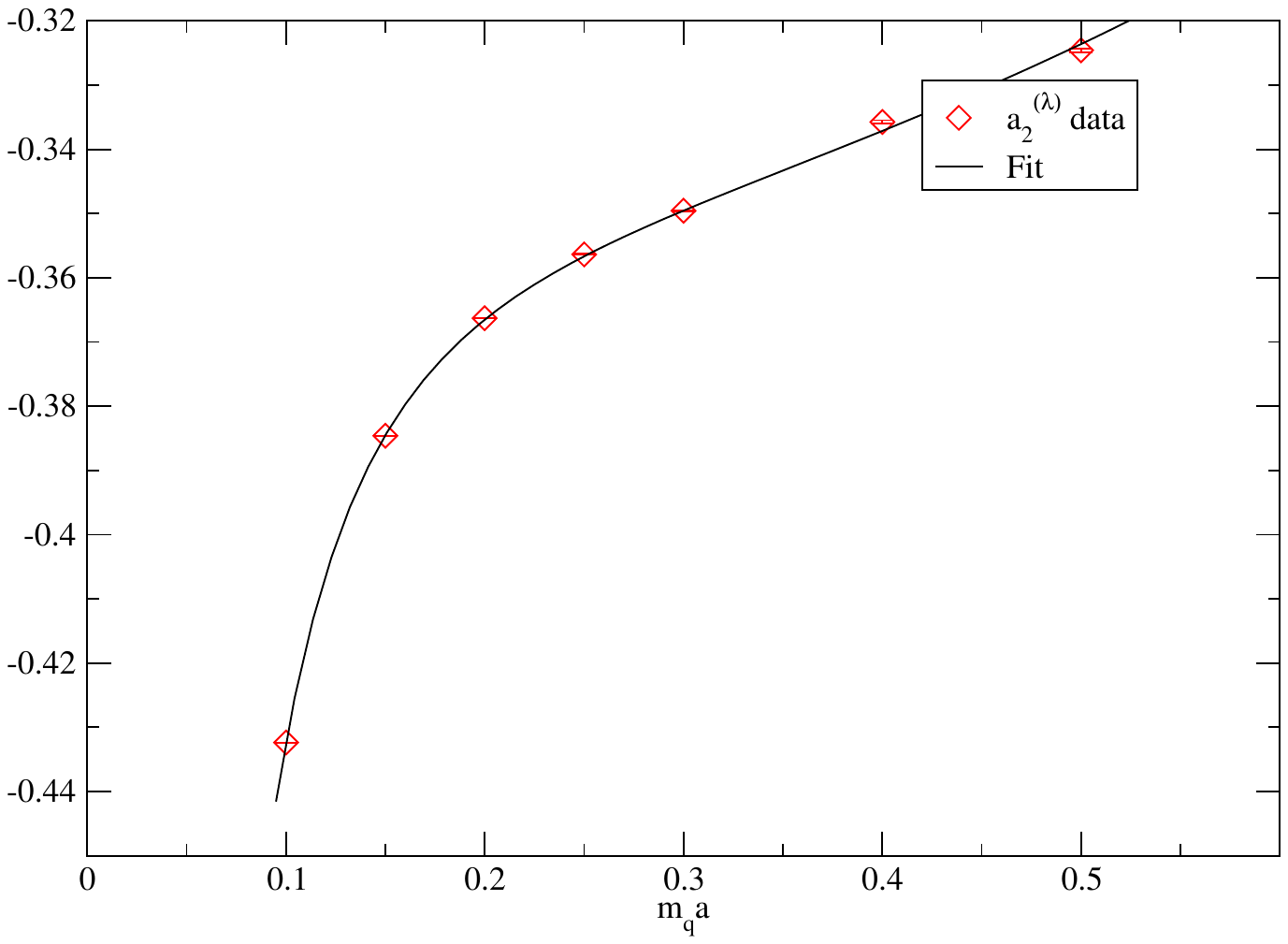}
\caption{\label{fig_2plots}
  Plots of $a_0^{(\lambda,1)}$ against $m_qa$ (left)
  and of $a_2^{(\lambda,1)}$ against $m_qa$ (right) with the fits
  shown for comparison.
  }
\end{figure*}

\subsection{Twisted boundary conditions}

We work on a four-dimensional Euclidean lattice of length $La$ in the
$x$ and $y$ directions and lengths $L_za,~L_ta$ in the $z$ and $t$ 
directions, respectively, where $a$ is the lattice spacing and $L,L_z,L_t$ are
even integers. In the following, we will employ twisted boundary conditions
\cite{'tHooft:1979uj}
for the same purpose and in essentially the same way as in
Refs.~\cite{Luscher:1985wf,Snippe:1997ru}.
The twisted boundary conditions we use for gluons and quarks are
applied to the $(x,y)$ directions and are given by ($\nu=x,y$)
\begin{align}
U_\mu(x+L\hat{\nu})  & =  \Omega_\nu U_\mu(x) \Omega_\nu^{-1}\;,
\nonumber \\
\Psi(x+L\hat{\nu})  & =  \Omega_\nu \Psi(x) \Omega_\nu^{-1}\;,
\end{align}
where the quark field $\Psi_{sc}(x)$ becomes a matrix in smell-colour
space
\cite{Parisi:1984cy}
by the introduction of a new SU($N$) quantum number ``smell''
in addition to the quark colour. In the $(z,t)$ directions, we apply
periodic boundary conditions.

These boundary conditions lead to a change in the Fourier expansion of
the fields:
\begin{align}
A_\mu(x) & = \frac{1}{N L^2 L_z L_t} \sum_{k} \Gamma_k e^{ikx}
\tilde{A}_\mu(k) 
\nonumber \\
\Psi_\alpha(x) & = \frac{1}{N L^2 L_z L_t} \sum_{p} \Gamma_p e^{ipx}
\tilde{\Psi}_\alpha(x)
\end{align}
where the matrices $\Gamma_k$ are given by (up to an
arbitrary phase, which may be chosen for convenience)
\begin{equation}
\Gamma_k = \Omega_1^{-n_2}\Omega_2^{n_1}
\end{equation}
and in the twisted $(x,y)$ directions the momentum sums are now over
\begin{equation}
p_\nu = m n_\nu,~~-\frac{NL}{2} < n_\nu \le \frac{NL}{2},~~\nu = (x,y)\;,
\end{equation}
where $m = \frac{2\pi}{N L a}$.
The zero modes ($n_x=n_y=0 \textrm{ mod } N$) are omitted from the sum
in the case of the gluons, but not the quarks.
The momentum sums for quark loops need to be divided by $N$ to remove
the redundant smell factor.

We may consider the continuum limit of the twisted theory
as a Kaluza-Klein theory in the infinite $(z,t)$ plane.
Denoting $\mathbf{n}=(n_x,n_y)$, the
stable particles in the $(z,t)$ continuum limit of this effective
theory are called the A mesons ($\mathbf{n}=(1,0)$ or
$\mathbf{n}=(0,1)$) with mass $m$ and the B mesons
($\mathbf{n}=(1,1)$) with mass $\sqrt{2}m$
\cite{Snippe:1997ru}.
%


\subsection{Small-mass expansions}

Even though we are ultimately interested in the radiative corrections
in the chiral limit, we cannot set $m_qa=0$ straightaway: the correct
way to approach the chiral limit is to maintain $m_q/m > C$ as we take
$m_qa \to 0$ and $ma \to 0$, where $C$ is a constant determined by the
requirement that a Wick rotation can be performed without encountering
a pinch singularity
\cite{Hao:2007iz}.

We therefore adopt the following procedure to extract the
$\mathcal{O}(a^2)$ lattice artifacts:
First, we expand some observable quantity $Q$ in powers
of $ma$ at fixed $m_qa$:
\begin{multline}\label{eqn:fit_in_ma}
Q(ma,m_qa)=a^{(Q)}_0(m_qa) + a^{(Q)}_2(m_qa) (ma)^2 + \\
\mathcal{O}\left((ma)^4,(ma)^4\log(ma)\right)
\end{multline}
where the coefficients in the expansion are all functions of $m_qa$.
There is no term at $\mathcal{O}\left((ma)^2\log(ma)\right)$ since the
gluon action is improved at tree-level to $\mathcal{O}(a^2)$
\cite{Snippe:1997ru}.
Then, we expand the coefficients $a^{(Q)}_0(m_qa)$ in powers of $m_qa$.

For $a^{(Q)}_0(m_qa)$ we have
\cite{Adams:2007gh}
\begin{equation}\label{eqn:fit0_in_mqa}
a^{(Q)}_0(m_qa)~=~b^{(Q)}_{0,0}\log(m_qa) + a^{(Q)}_{0,0}\;.
\end{equation}
Since we expect a well-defined continuum limit, $a^{(Q)}_0(m_qa)$
cannot contain any negative powers of $m_qa$, but, depending on
the quantity $Q$, it may contain logarithms; $b^{(Q)}_{0,0}$ is the
anomalous dimension associated
with $Q$, and can be determined by a continuum calculation.
There can be no terms in $(m_qa)^{2n}$ for $n>0$ since there is no
counterterm in the gluon action that can compensate for a scaling
violation of this kind.

For $a^{(Q)}_2(m_qa)$ we find
\begin{multline}\label{eqn:fit2_in_mqa}
a^{(Q)}_2(m_qa)~=~\frac{a^{(Q)}_{2,-2}}{(m_qa)^2} + a^{(Q)}_{2,0}
+ \\
\left(a^{(Q)}_{2,2} + b^{(Q)}_{2,2}\log(m_qa)\right)(m_qa)^2 +
   \mathcal{O}\left((m_qa)^4\right)\;.
\end{multline}
After multiplication by $(ma)^2$, the $(m_qa)^{-2}$ contribution gives
rise to a continuum contribution to $Q$, and $a^{(Q)}_{2,-2}$ is
calculable in continuum perturbation theory. There can be no term in
$(m_qa)^{-2}\log(m_qa)$ since this would be a volume-dependent further
contribution to the anomalous dimension of $Q$, and there can be no
term in $\log(m_qa)$ since the action is tree-level $\mathcal{O}(a^2)$
improved
\cite{Symanzik:1983dc}.
A rigorous proof of Eqn.~(\ref{eqn:fit2_in_mqa})
along the lines of Ref.~\cite{Adams:2007gh}
would, of course, be welcome.

In the chiral limit $m_q\to 0$, the term $w_i$ that appears on the
right-hand side of Eqn.
(\ref{eqn:impcond_generic})
is $a^{(Q)}_{2,0}$.


\subsection{Twisted spectral quantities}

The simplest spectral quantity that can be chosen within the framework
of the twisted boundary conditions outlined above is the
(renormalised) mass of the A meson. The one-loop correction to the A
meson mass is given by
\cite{Snippe:1997ru}
\begin{equation}\label{eqn:mA}
m_A^{(1)} = - Z_0(\mathbf{k})
              \left.\frac{\pi_{11}^{(1)}(k)}{2 m_A^{(0)}}\right|
              _{k=(i m_A^{(0)},0,m,0)}
\end{equation}
where $Z_0(\mathbf{k})=1+\mathcal{O}\left((ma)^4\right)$ is the
residue of the pole of the tree-level gluon propagator at spatial
momentum $\mathbf{k}$, and $m_A^{(0)}$ is defined so that the momentum
$k$ is on-shell. 

Gauge invariance implies
\cite{Hao:2007iz}
\begin{align}
a^{(m_A,1)}_{2,-2} & = 0 \; , 
\nonumber \\
a^{(m_A,1)}_0(m_qa) & = 0 \; .
\end{align}
The $\mathcal{O}\left(\alpha_s (ma)^2\right)$ contribution from
improvement of the action is given by
\cite{Snippe:1997ru}
\begin{equation}
\Delta_\textrm{imp} \frac{m_A^{(1)}}{m} = 
- ( c_1^{(1)} - c_2^{(1)} ) (ma)^2 + \mathcal{O}\left((ma)^4\right)\; ,
\end{equation}
leading to the improvement condition
\begin{equation}
c_1^{(1)} - c_2^{(1)} = a_{2,0}^{(m_A,1)} \; .
\label{eqn_imp_cond_1}
\end{equation}
The next simplest independent spectral quantity is the
scattering amplitude for A mesons at B meson threshold, which can be
described by an effective $AAB$ meson coupling constant $\lambda$
\cite{Luscher:1985zq}:
\begin{equation}
\label{eqn:def_of_lambda}
\lambda =
g_0 \sqrt{ Z(\mathbf{k}) Z(\mathbf{p}) Z(\mathbf{q}) }
\; e_j \; \Gamma^{1,2,j}(k,p,q)
\end{equation}
where a twist factor of $\frac{i}{N}\mathrm{Tr}([\Gamma_k,\Gamma_p]\Gamma_q)$
has been factored out from from both sides,
and the momenta and polarisations of the incoming particles are
(with $r>0$ defined such that $E(\mathbf{q})=0$)
\begin{equation}\label{eqn:momenta}
\begin{array}{ll}
k = (iE(\mathbf{k}),\mathbf{k}),  & \mathbf{k} = (0,m,ir) \\
p = (-iE(\mathbf{p}),\mathbf{p}),   \quad      & \mathbf{p} = (m,0,ir) \\
q = (0,\mathbf{q}),                      &  \mathbf{q} = (-m,-m,-2ir) \\
e = (0,1,-1,0) \\
\end{array}
\end{equation}
We expand Eqn.~(\ref{eqn:def_of_lambda}) perturbatively to one-loop
order and find (up to $\mathcal{O}((ma)^4)$ corrections)
\begin{multline}
\frac{\lambda^{(1)}}{m} =
\left( 1 - \frac{1}{24} m^2 \right)\frac{\Gamma^{(1)}}{m}
- \frac{4}{k_0} \frac{d}{dk_0}
           \left.\pi_{11}^{(1)}(k)\right|_{k_0=iE(\mathbf{k})} \\
-  \left( 1 - \frac{1}{12} m^2 \right) \frac{d^2}{dq_0^2}
           \left. \left( e^i e^j \pi_{ij}^{(1)}(q) \right) \right|_{q_0=0}
\end{multline}
where $\Gamma^{(1)}$ is the one-particle irreducible three-point function
at one loop. The derivatives of the Feynman diagrams contributing to the
self-energy are computed analytically using automatic differentiation
\cite{vonHippel:2005dh,vonHippel:unpub}.
Continuum calculations of the anomalous dimension and infrared
divergence give
\begin{align}
b^{(\lambda,1)}_{0,0} & = -\frac{N_f}{3\pi^2}g^2\;,
\nonumber \\
a^{(\lambda,1)}_{2,-2} & = -\frac{N_f}{120\pi^2}g^2\;.
\label{eqn:anom_dir_lambda}
\end{align} 
The improvement contribution to $\lambda$ is
\cite{Snippe:1997ru}
\begin{equation}
\Delta_\textrm{imp} \frac{\lambda^{(1)}}{m} = 4 (9 c_1^{(1)} - 
7 c_2^{(1)}) (ma)^2
   + \mathcal{O}\left((ma)^4\right)\;,
\end{equation}
leading to the improvement condition
\begin{equation}
4\left(9c_1^{(1)} - 7c_2^{(1)}\right) = -a_{2,0}^{(\lambda,1)} \; .
\label{eqn_imp_cond_2}
\end{equation}
%


\section{Results}

To extract the improvement coefficients from our diagrammatic
calculations, we compute the diagrams for a number of different values
of both $L$ and $m_q$ with $N_f=1$, $N=3$. At each value of $m_q$, we
then perform a fit in $ma$ of the form given in
Eqn.~(\ref{eqn:fit_in_ma}) to extract the coefficients
$a_n^{(Q,1)}(m_qa)$ for $n=0,2$. Our fits confirm that
$a_0^{(m_A,1)}(m_qa)=0$; an example is shown in Fig.~\ref{fig_1plot}

Performing a fit of the form in
Eqns.~(\ref{eqn:fit0_in_mqa},\ref{eqn:fit2_in_mqa}) respectively on
these coefficients, we are able to extract the analytically-known
coefficients with high accuracy along with the required $(ma)^2$
contributions, as shown in Fig.~\ref{fig_2plots}.

Our results for the fermionic contributions are
\begin{align}
a_{2,0}^{(m_A,1)} & = 0.00942(3) \; ,
\nonumber\\
a_{2,0}^{(\lambda,1)} & = -0.352(2) \; .
\end{align}
Equating these results with the $w_i$ of
Eqn.~(\ref{eqn:impcond_generic}), we can solve
Eqns.~(\ref{eqn_imp_cond_1},\ref{eqn_imp_cond_2}) for $c_i^{(1)}$ to
obtain
\begin{align}
c_1^{(1)} & = -0.025218(4) + 0.0110(3) N_f 
\nonumber \\
c_2^{(1)} & = -0.004418(4) + 0.0016(3) N_f
\nonumber \\
\Rightarrow \quad
c_0^{(1)} & = 0.237088 (46) - 0.1008(34) N_f
\label{eqn_shift_c}
\end{align}
where the quenched ($N_f=0$) results are taken from
Ref.~\cite{Snippe:1997ru}
and we have propagated the errors by quadrature into $c_0^{(1)}$.

\begin{figure}[t]
\begin{center}
\includegraphics[angle=-90,width=2in]{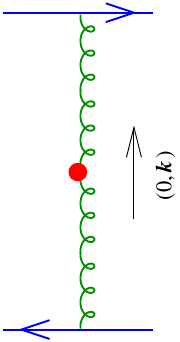}
\end{center}
\caption{\label{fig_feynman} The one-loop counterterm contribution to 
the static potential.}
\end{figure}

\subsection{Effect on gauge action couplings}

The MILC and UKQCD Collaborations use a ``tadpole improved'' version of
Eqn.~(\ref{eqn:lw_action}), dividing each gauge link by a factor
$u_0$. In addition, a factor of $c_0/u_0^4$ is subsumed into the gauge
coupling $\beta_0 = 6c_0/(g^2 u_0^4)$ that multiplies the plaquette
term $P_0$
\cite{Alford:1995hw}. 
The couplings multiplying the ``planar rectangles'' $P_1$ and
``parallelograms'' $P_2$ are
\cite{Alford:1995hw,Hao:2007iz}
\begin{align}
\beta_1 & = -\frac{\beta_0}{20u_0^2}\left[1-\left(\frac{12\pi}{5}c_0^{(1)}+
 48\pi c_1^{(1)}+2u_0^{(1)}\right)\alpha_s\right]\;,
\nonumber\\
\beta_2 & = \frac{12\pi\beta_0}{5u_0^2}c_2^{(1)}\alpha_s\;,
\end{align}
(with factors of $4\pi$ coming from converting from $g^2$ to
$\alpha_s$). The quenched radiative contributions have been analyzed
in
\cite{Alford:1995hw}
and so we may write
\begin{align}
\beta_1 & = -\frac{\beta_0}{20u_0^2}\left[1+0.4805\alpha_s-
\left(\frac{12\pi}{5}c_{0,f}^{(1)}+
48\pi c_{1,f}^{(1)}\right)\alpha_s\right]\;,
\nonumber\\
\beta_2 & = -\frac{\beta_0}{u_0^2}\left(0.033\alpha_s-
\frac{12\pi}{5}c_{2,f}^{(1)}\alpha_s\right)\;,
\end{align}
where now all the one-loop coefficients $c_{i,f}^{(1)}$ contain only
quark loop contributions.

Plugging in the numbers for the HISQ action obtained in this work we find
\begin{align}
\beta_1 & = -\frac{\beta_0}{20u_0^2}\left[1+0.4805\alpha_s
-0.899(52)N_f\alpha_s\right]\;,
\nonumber\\
\beta_2 & = -\frac{\beta_0}{u_0^2}\left[0.033\alpha_s
-0.0121(23)N_f\alpha_s\right] \; .
\label{eqn_shift_beta}
\end{align}
The full coefficient $c_0$ has here been absorbed into the gauge
coupling, so the coefficient multiplying the plaquette $P_0$ is simply
$\beta_p = \beta_0$.  For the HISQ action, the fermionic contribution
to $c_0$ (i.e. $c_{0,f}^{(1)}$ in Eqn.~(\ref{eqn_shift_c})) is large
and sizeable shifts will be needed in $\beta_0$ to maintain a constant
$g^2$ (or lattice spacing) as $N_f$ is changed from $0$ (quenched) to
$N_f = 3$ or~$4$. Whilst this is not a problem in itself, it does make
it more difficult to intuitively relate values of $\beta_0$ to the
lattice spacing.

A more sensible choice is to absorb just the tree-level portion
$c_0^{(0)}=\frac{5}{3}$ into the gauge coupling.  The overall gauge
coupling is simply $\beta_0^\prime = 10/(g^2 u_0^4)$.  Using primes to
denote couplings in this scheme, the coupling multiplying the
plaquette in the action is now
\begin{align}
\beta_p^\prime  & = \beta_0^\prime 
\left[ 1 + \frac{4 \pi c_0^{(1)}}{c_0^{(0)}} \alpha_s  \right]
\nonumber \\
& = \beta_0^\prime 
\left[ 1 + 1.7876 \alpha_s - 0.760 (26) N_f \alpha_s \right] \; .
\end{align}
The remaining couplings in this scheme are
\begin{align}
\beta_1^\prime & = -\frac{\beta_0^\prime}{20u_0^2}\left[1-
\left(48\pi c_1^{(1)}+2u_0^{(1)}\right)\alpha_s\right]
\nonumber\\
& = -\frac{\beta_0^\prime}{20u_0^2}\left[1+2.2681\alpha_s-
\left(
48\pi c_{1,f}^{(1)}\right)\alpha_s\right]
\nonumber\\
& = -\frac{\beta_0^\prime}{20u_0^2}\left[1+2.2681 \alpha_s - 
1.659 (46)
N_f\alpha_s\right]\;,
\nonumber\\
\beta_2^\prime & = \frac{12\pi\beta_0^\prime}{5u_0^2}c_2^{(1)}\alpha_s
\nonumber \\
& = -\frac{\beta_0^\prime}{u_0^2}\left[0.033\alpha_s
-0.0121(23)N_f\alpha_s\right] \; .
\end{align}
The factors multiplying the gauge coupling in $\beta_2$ and
$\beta_2^\prime$ are the same as this term is already
$\mathcal{O}(\alpha_s)$.

\section{Radiative improvement and the static potential}
\label{sec_milc_pot}

\begin{figure}[t]
\begin{center}
\includegraphics[width=3in,clip]{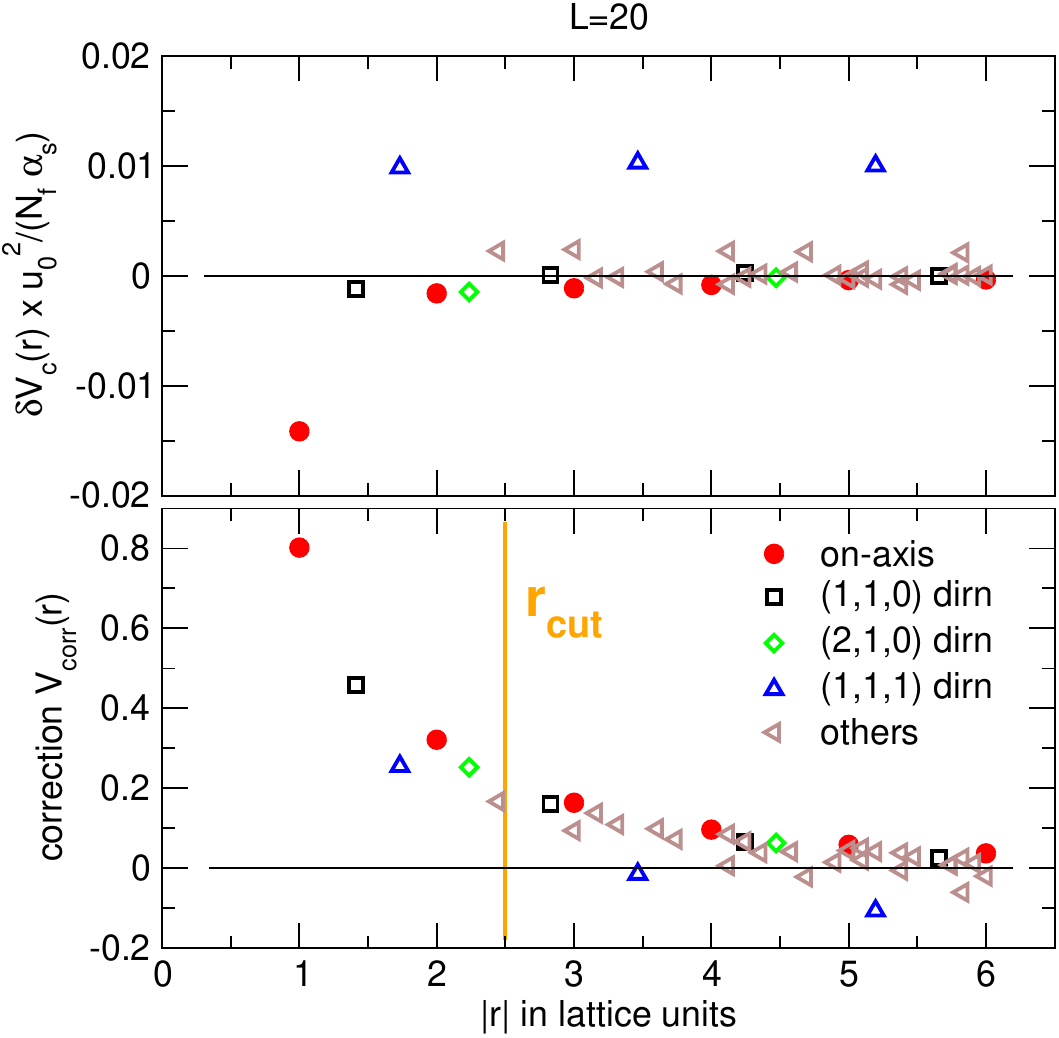}
\end{center}
\caption{\label{fig_deltaV} The perturbative correction 
(top panel) and the correction for lack of rotational 
invariance (bottom panel).}
\end{figure}

In this section, we seek to understand what effect the omission of the
asqtad $\mathcal{O}(N_f \alpha_s a^2)$ corrections to gauge action
will have on physical observables measured in existing
non-perturbative Monte Carlo lattice simulations.

As discussed above, the then-unknown $\mathcal{O}(N_f \alpha_s a^2)$
contributions to $c_i$ were omitted from the current generation of
three-flavour dynamical asqtad simulations.  Perturbatively, this
omission will lead to an imperfect cancellation of discretisation
effects and a residual breaking of rotational symmetry in the static
quark potential. Similar effects are expected to be seen in the
numerical simulation results, and hence in the determinations of the
scale-setting parameters $\rone$,~$\rzero$ derived from the static
potential.

Here we use one-loop perturbation theory to calculate the effect of
the missing $\mathcal{O}(N_f \alpha_s a^2)$ terms on the static
potential and particularly on the scale-setting parameters $\rone$,
$\rzero$. The rationale for this and alternative approaches are
discussed in Secs.~\ref{sec_intro} and~\ref{sec_disco}.

\subsection{The correction to the static potential}

Including the missing $\mathcal{O}(N_f \alpha_s a^2)$ corrections to
the gauge action would change the lattice static-quark potential
$V(\br)$ measured by the MILC Collaboration by an amount $\delta
V(\br)$. Here we estimate this change using one-loop perturbation
theory.

We do this by computing the spatial Fourier transform (for zero
temporal momentum) of $G_{0 \mu}(k) \Delta_{\mu \nu}(k) G_{\nu 0}(k)$,
where $G_{\mu \nu}(k)$ is the tree-level Symanzik improved gluon
propagator at momentum $k = (0,\bk)$, as shown in
Fig.~\ref{fig_feynman}. $\Delta_{\mu \nu}(k)$ is the
$\mathcal{O}(g^2)$ insertion into the gluon propagator arising from
the perturbative expansion of the $\mathcal{O}(N_f\alpha_s a^2)$
corrections to the gauge action in Eqn.~(\ref{eqn:lw_action}) with
appropriate asqtad $c_{i,f}^{(1)}$ couplings
\cite{Hao:2007iz}.
Again, Feynman rules are derived using the \textsc{HiPPy} package
\cite{Hart:2004bd,Hart:prog}.

In all cases, the spatial Fourier transforms are carried out for a
finite, periodic lattice of spatial volume $L^3$. 

The gauge action has $c_3=0$, but the static quark counterterm
proportional to $c_4$ is also omitted. This will also affect the
success of the radiative improvement, but we do not consider its
effect in this paper.

The result for $L=20$ is shown in the upper panel of
Fig.~\ref{fig_deltaV}. We expect the \textit{corrected} lattice
potential to be rotationally symmetric at $\mathcal{O}(\alpha_s a^2)$,
so the lack of rotational invariance in the upper panel of
Fig.~\ref{fig_deltaV} is indicative of an equal and opposite breaking
of rotational symmetry in the potential measured on ensembles that
omit the $\mathcal{O}(N_f \alpha_s a^2)$ radiative corrections.

\subsection{The effect of the correction}

To set the scale, the MILC Collaboration measure the static potential
for a variety of on- and off-axis spatial separations: $\{ V(\br_i)
\}$ with associated statistical errors $\{ \sigma(\br_i) \}$. A least
squares fit is performed using the fit function
\cite{Bernard:2000gd,Bernard:2001av,Aubin:2004wf}
\begin{eqnarray}
V_{\text{fit}}(\br) & = & V_{\text{cont}}(r) + b_3 V_{\text{corr}}(\br)
\equiv \sum_{j=0}^3 b_j f_j(\br)\;,
\nonumber \\
V_{\text{cont}}(r) & = & b_0 - \frac{b_1}{r} + b_2 \; r \; ,
\nonumber \\
V_{\text{corr}}(\br) & = & 
\begin{cases}
V_{\text{free}}(\br) - \frac{1}{r} & r < r_{\text{cut}} \\
0 & \mbox{otherwise}
\end{cases}
\nonumber
\end{eqnarray}
which defines basis functions $f_j(\br)$ with a vector of fit
parameters $\bob$. $V_{\text{corr}}$ aims to account for the lack of
rotational invariance at small $r \equiv |\br|$, with
$V_{\text{free}}$ the finite-sized lattice estimate for $1/r$ from the
Fourier transform of the (free) Symanzik gluon Coulomb propagator. We
show this for $L=20$ in the lower panel of Fig.~\ref{fig_deltaV}.

In more detail, the fit minimises the least-squared
function
\ben
L(\bob) = \sum_{i}\;\frac{\bb V(\br_i) - 
V_{\text{fit}}(\br_i)\eb^2}{\sigma(\br_i)^2} \; .
\een
We define the (weighted) average of operator $A(\br)$ over a set of
measured $\br_i$ as
\ben
\la A \ra = \left. 
\sum_i^N \frac{A(\br_i)}{\sigma(\br_i)^2} 
\; \middle/ \; \sum_i^N \frac{1}{\sigma(\br_i)^2}
\right. \; .
\een
The result of the least squares fitting is a vector of best-fit
parameters $\bob$ that obeys the linear equation
\ben
M_{jk} \; b_k = X_j \quad \Rightarrow \quad \bob = M^{-1} \underline{X}
\een
where
\ben
M_{jk} = \la f_j f_k \ra, \quad X_j = \la f_j V \ra \; .
\een

Having done this, the lattice scale is set from the analytic
derivative of the $V_{\text{cont}}(r)$ function:
\ben
\left. \hat{r}_n^2\frac{dV_{\text{cont}}(r)}{dr}\right|_{\hat{r}_n} = C_n \quad
\Rightarrow \quad \hat{r}_n = \sqrt{\frac{C_n-b_1}{b_2}} \; .
\label{r1}
\een 
We use ``hats'' here to stress that the scale parameters are measured
in dimensionless lattice units.

Two scales are commonly used: $\rone$ from $C_1=1$ (physical value
$r_1 = 0.317$~fm
\cite{Aubin:2004wf}) 
and $\rzero$ from $C_0=1.65$ (physical
value $r_0 = 0.462$~fm
\cite{Aubin:2004wf}). 
In general $\rone$ is preferred as the statistical errors on the
static potential are smaller at shorter distances.

\subsection{The corrected fits}

Having calculated the $\mathcal{O}(N_f \alpha_s a^2)$ corrections to
the static potential, we can now calculate the effect of including
$\delta V$ on the best fit parameters $b_j$, assuming that we carry
out exactly the same fitting procedure as before.

We would now minimise
\ben
L^\prime(\bob) = \sum_i \frac{\bb V(\br_i) + \alpha_s \delta V(\br_i) - 
V_{\text{fit}}(\br_i)\eb^2}{\sigma(\br_i)^2}\;.
\een
Given that $\bob$ minimises $L(\bob)$, we assert that
$\bob^\prime=\bob+\alpha_s \delta \bob$ minimises $L^\prime(\bob)$, with
\ben
\delta \bob = M^{-1} \delta \underline{X}
\een
and $\delta X_j = \la f_j \; \delta V \ra$.

After finding $\delta\bob$ we can deduce the associated change in
$\rone$. We can either define this as $\delta \rone = \rone(\bob + \alpha_s
\delta \bob) - \rone(\bob)$ or, using a Taylor expansion of
Eqn.~(\ref{r1}),
\ben
\frac{\delta \rone}{\rone} = -\frac{1}{2}\bb \frac{\delta b_1}{1-b_1} + 
\frac{\delta b_2}{b_2} \eb\;.
\een
The two methods give almost identical results.

\subsection{Results}

We looked at a range of ensembles listed in Table~\ref{tab_results},
using published values of $u_0$ to infer the strong coupling constant
in the same way as MILC
\cite{Orginos:1998ue}:
\ben
\alpha_s = -\frac{4 \log u_0}{3.0684} \; .
\een
To estimate the effect the fermionic corrections would have on the
scale setting parameters as measured by the MILC Collaboration, we
adopt the same fitting function and we use the same fit range
$\sqrt{5} \le r \le 7$ and $r_{\text{cut}} = 3$ for the ``fine''
lattices and $\sqrt{2} \le r \le 6$ with $r_{\text{cut}} = 2.5$ for
the ``coarse'' and ``very coarse'' ensembles.

We infer $b_1$ and $b_2$ from published values for $\rone$ and
$\rzero$ on given ensembles
\cite{Bernard:2001av,priv:comm}:
\ben
b_2 = \frac{1.65 - 1}{\rzero^2 - \rone^2}, \quad
b_1 = 1 - b_2 \; \rone^2 \; .
\een
For instance, on the $\beta=6.76$, $m_u/m_s = 0.01/0.05$ coarse
ensemble $\rone = 2.60$
(Ref.~\cite{Bernard:2001av}), 
$\rzero = 3.76$ 
(Ref.~\cite{priv:comm})
giving $b_1 = 0.406$, $b_2=0.088$. We then find 
\ben
\delta \rone/\rone =
-0.65 \%, \quad \delta \rzero/\rzero = -0.11 \%.
\een
The shift in $\rone$ is larger because $\rone$ is smaller than
$\rzero$ and $\delta V$ is short-ranged. On the fine lattices,
$\rone$ in lattice units is comparable to $\rzero$ on the coarse
lattices. The shift is therefore small. Results for other ensembles
are given in Table~\ref{tab_results}

We have looked at various scenarios, e.g. different choices for the
fitted range of $\{\br_i\}$ and constraining some fit parameters to
zero. Whilst the precise shifts do vary, the scale (and sign) of the
shifts remain stable under such variations.

\begin{table*}[t]
\caption{MILC simulation parameters and shifts in scale setting
parameters induced by \textit{omission} of fermionic radiative
corrections to the gluonic action. Smoothed $\rone$ values are from
Ref.~\cite{Follana:2007uv,Aubin:2004wf}.  $\rzero$ values are then
inferred from the ratios $\rzero/\rone$ given in
Ref.~\cite{Aubin:2004wf}. We have estimated $u_0$ for the very coarse
ensemble. Lattice spacings are quoted as approximate guides; precise
values may be inferred from setting $r_1 = 0.317$~fm.}
\label{tab_results}

\begin{ruledtabular}
\begin{tabular}{ccccccccc}
Label & $a$/fm (approx) & $L^3\times T$ &
Sea quark & $\rone$ & $\rzero$ & $u_0$ & $\delta \rone/\rone$  & 
$\delta \rzero/\rzero$ \\
& & & masses $m_l/m_s$ & & & & (in \%) & (in \%) \\
\hline
\textit{very coarse} & $0.18$ & $16^3 \times 48$ &
0.082/0.082 & 1.805~(10) & 2.622~(28) & 0.8585 & $-1.11$ & $-0.40$ \\
\hline
\textit{coarse} & $0.12$  & $20^3 \times 64$ &
0.02/0.05 & 2.650~(8) & 3.828~(15) & 0.8688 & $-0.63$ & $-0.11$ \\
& & &
0.01/0.05 & 2.610~(12) & 3.774~(20) & 0.8677 & $-0.65$ & $-0.11$ \\
& & $24^3 \times 64$ & 
0.005/0.05 & 2.632~(13) & 3.834~(25) & 0.8678 & $-0.64$ & $-0.10$ \\
\hline
\textit{fine} & $0.09$ & $28^3 \times 96$ &
0.0124/0.031 & 3.711~(13) & 5.398~(28) & 0.8788 & 0.01 & 0.00 \\
& & &
0.0062/0.031 & 3.684~(12) & 5.384~(27) & 0.8782 & 0.01 & 0.00 \\
\end{tabular}
\end{ruledtabular}

\end{table*}
\section{Discussion}
\label{sec_disco}

Radiatively improved gluon actions are used in lattice simulations to
give greater control over discretisation effects and to reduce the
uncertainty in continuum-extrapolated quantities. A typical example is
the use of the L\"uscher-Weisz action in improved staggered
simulations by the MILC and UKQCD Collaborations.

We note that current unquenched simulations employing lattice quark
formulations other than improved staggered, such as domain wall or
improved Wilson clover, generally do not use a radiatively improved
action for the gluons; hence a calculation of the effects of fermion
loops on the gluonic action is currently neither necessary nor useful
for those simulations, but could readily be performed if and when
simulations using such quark actions together with the L\"uscher-Weisz
action will be undertaken.

Simulations employing staggered quarks rely on the validity of the
``fourth root trick'', which has not yet been rigorously established.
The purpose of this paper is not to engage in the debate about the
validity of this procedure, but merely point out that simulations
using improved staggered quarks have produced results in excellent
agreement with experiment so far. While we cannot completely discard
the possibility that the observed scaling violation in $\Phi_B$ might
be an indication of some more fundamental problem, we believe that our
explanation for this scaling violation is more likely in the light of
existing evidence. In particular, we are able to replicate both the
sign and the rough magnitude of the observed effect by a perturbative
calculation.

The shift in the radiative corrections due to the HISQ fermions in
Eqns.~(\ref{eqn_shift_c},\ref{eqn_shift_beta}) is surprisingly large,
even compared to the coefficients for asqtad fermions
\cite{Hao:2007iz}.
At first sight, this may seem like a surprise, since HISQ is supposed
to be the more highly-improved action. However, HISQ is designed to
suppress taste-changing interactions coming from low momentum
quark/high momentum gluon couplings, but the gluonic improvement
coefficients come from high momentum quark/low momentum gluon couplings,
for whose suppression the HISQ action is not tuned.

One consequence is that if the coefficient $c_0$ is subsumed into the
leading factor of $\beta_0$, we expect to see large $N_f$ dependent
shifts in the value of $\beta_0$ at fixed $g^2$, and we also give
results for an alternate scheme where only the tree level part of
$c_0$ is included in the overall gauge coupling.

The radiative corrections in the L\"uscher-Weisz action used by MILC
in the asqtad simulations, however, omit the contribution from
dynamical sea quarks. This contribution has recently been calculated
at one-loop for $N_f$ (massless) flavours of asqtad improved staggered
fermions. The results are dramatic, leading to sign reversals in some
of the radiative coefficients.

It is therefore conceivable that the omission of the $\mathcal{O}(N_f
\alpha_s a^2)$ corrections leads to increased scaling violations in
results from dynamical simulations when compared to quenched data.

To properly establish whether this is the case would require a new set
of dynamical Monte Carlo simulations, which is well beyond the scope
of this study. An alternative is to attempt a reweighting of the
existing ensembles using factors $e^{-\delta S}$ based on the
$\mathcal{O}(N_f \alpha_s a^2)$ counterterms. Such calculations
notoriously suffer from a very poor overlap between the importance
samplings of the original and reweighted ensembles for even minor
changes in the action. This leads to very large statistical errors
which will obscure any sought-for effect.

We have therefore used instead one-loop lattice perturbation theory to
estimate the effect of the $\mathcal{O}(N_f \alpha_s a^2)$ corrections
on the static potential and the shifts in the scale-setting parameters
$\rone$ and $\rzero$ arising from the \textit{omission} of the
fermionic radiative corrections for typical values of the simulations
with $2+1$ dynamical asqtad flavours.

On \textit{fine} ($a \simeq 0.09$~fm) lattices, the shifts in $\rone$
and $\rzero$ are negligible (less than $0.1\%$) and will be at least
as small on \textit{superfine} lattices with $a \simeq 0.06$~fm. On
\textit{coarse} lattices ($a \simeq 0.12$~fm), omission of the
corrections leads to $\rone$ being $0.6\%$ too large, with $\rzero$
unaffected. On \textit{very coarse} lattices ($a \simeq 0.18$~fm),
$\rone$ is $1.1\%$ too large and $\rzero$ $0.4\%$ too large.

Overall, then, the omission of the $\mathcal{O}(N_f \alpha_s a^2)$
leads to an underestimate of the lattice spacing on coarser lattices
as defined using $\rone$.  Whilst numerically small, this effect is
comparable to the statistical errors on a number of quantities and
therefore would lead to a measurable increase in the statistical
uncertainty of continuum-extrapolated lattice QCD predictions.

Higher loop and non-perturbative effects will almost certainly change
the exact value of the shift in $\rone$, but are unlikely to alter our
main conclusion: that the effect is measurable.

Putting aside the static potential, an alternative approach to fixing
the lattice spacing is to use the $2S-1S$ mass splitting of $\Upsilon$
states. We have seen that that the correction $\delta V(\br)$ is
negative, so including the $\mathcal{O}(N_f \alpha_s a^2)$ radiative
corrections would decrease slightly both (lattice) $\Upsilon$
masses. Because $\delta V(\br)$ is short-ranged, the effect on the
$1S$ state will be larger than on the $2S$ state since the $1S$
wavefunction is larger at small $r$. The lattice mass splitting and
thus the derived value of $a$ will increase, and thus $\Phi_B$ will
get slightly smaller in such physical units on coarse lattices. To
reliably deduce this fact from the $\Upsilon$ mass gap, however, we
need to include the effect of the contact term,
Eqn.~(\ref{eqn_contact}), which we do not yet know.

\section*{Acknowledgments}

We thank C.T.H. Davies, U.M. Heller, G.P. Lepage and D. Toussaint for
useful conversations and comments. A.H.~thanks the U.K.~Royal Society
for financial support. G.M.v.H.~was supported by the Deutsche
Forschungsgemeinschaft in the SFB/TR 09. This work has made use of the
resources provided by: the Darwin Supercomputer of the University of
Cambridge High Performance Computing Service
(\url{http://www.hpc.cam.ac.uk}), provided by Dell Inc.~using
Strategic Research Infrastructure Funding from the Higher Education
Funding Council for England; the Edinburgh Compute and Data Facility
(\url{http://www.ecdf.ed.ac.uk}), which is partially supported by the
eDIKT initiative (\url{http://www.edikt.org.uk}); the Fermilab Lattice
Gauge Theory Computational Facility. The University of Edinburgh is
supported in part by the Scottish Universities Physics Alliance
(SUPA).


\begin{thebibliography}{10}

\bibitem{Davies:2003ik}
HPQCD, C.~T.~H. Davies {\em et~al.},
\newblock Phys. Rev. Lett. {\bf 92}, 022001 (2004), [hep-lat/0304004].

\bibitem{Orginos:1999cr}
MILC, K.~Orginos, D.~Toussaint and R.~L. Sugar,
\newblock Phys. Rev. {\bf D60}, 054503 (1999), [hep-lat/9903032].

\bibitem{Follana:2006rc}
HPQCD, E.~Follana {\em et~al.},
\newblock Phys. Rev. {\bf D75}, 054502 (2007), [hep-lat/0610092].

\bibitem{Wong:2007uz}
K.~Y. Wong and R.~M. Woloshyn,
\newblock PoS {\bf LAT2007}, 047 (2007), [0710.0737].

\bibitem{Bazavov:2009jc}
MILC, A.~Bazavov {\em et~al.},
\newblock PoS {\bf LAT2008}, 033 (2008), [0903.0874].

\bibitem{Hao:2007iz}
Z.~Hao, G.~M. von Hippel, R.~R. Horgan, Q.~J. Mason and H.~D. Trottier,
\newblock Phys. Rev. {\bf D76}, 034507 (2007), [0705.4660].

\bibitem{Hart:2008zi}
A.~Hart, G.~M. von Hippel and R.~R. Horgan,
\newblock PoS {\bf LAT2008}, 046 (2008), [0808.1791].

\bibitem{Gregory:2008mn}
UKQCD, E.~B. Gregory, A.~C. Irving, C.~McNeile and C.~M. Richards,
\newblock 0810.0136.

\bibitem{Gray:2005ad}
HPQCD, A.~Gray {\em et~al.},
\newblock Phys. Rev. Lett. {\bf 95}, 212001 (2005), [hep-lat/0507015].

\bibitem{Davies:comm}
C.~Davies,
\newblock private communication.

\bibitem{WeiWoh:1}
P.~Weisz and R.~Wohlert,
\newblock Nucl. Phys. B {\bf 236}, 397 (1984).

\bibitem{Snippe:1997ru}
J.~R. Snippe,
\newblock Nucl. Phys. {\bf B498}, 347 (1997), [hep-lat/9701002].

\bibitem{Luscher:1985zq}
M.~L\"{u}scher and P.~Weisz,
\newblock Phys. Lett. {\bf B158}, 250 (1985).

\bibitem{Wei:1}
P.~Weisz,
\newblock Nucl. Phys. B {\bf 212}, 1 (1983).

\bibitem{Alford:1995hw}
M.~G. Alford, W.~Dimm, G.~P. Lepage, G.~Hockney and P.~B. Mackenzie,
\newblock Phys. Lett. {\bf B361}, 87 (1995), [hep-lat/9507010].

\bibitem{Lepage:1996jw}
G.~P. Lepage,
\newblock hep-lat/9607076.

\bibitem{Luscher:1985wf}
M.~L\"{u}scher and P.~Weisz,
\newblock Nucl. Phys. {\bf B266}, 309 (1986).

\bibitem{Luscher:1984xn}
M.~L\"{u}scher and P.~Weisz,
\newblock Commun. Math. Phys. {\bf 97}, 59 (1985).

\bibitem{Hart:2004bd}
A.~Hart, G.~M. von Hippel, R.~R. Horgan and L.~C. Storoni,
\newblock J. Comput. Phys. {\bf 209}, 340 (2005), [hep-lat/0411026].

\bibitem{Hart:prog}
A.~Hart, G.~M. von Hippel and R.~R. Horgan,
\newblock Comput. Phys. Commun., in preparation.

\bibitem{Drummond:2002kp}
I.~T. Drummond, A.~Hart, R.~R. Horgan and L.~C. Storoni,
\newblock Nucl. Phys. Proc. Suppl. {\bf 119}, 470 (2003), [hep-lat/0209130].

\bibitem{Nobes:2001tf}
M.~A. Nobes, H.~D. Trottier, G.~P. Lepage and Q.~Mason,
\newblock Nucl. Phys. Proc. Suppl. {\bf 106}, 838 (2002), [hep-lat/0110051].

\bibitem{Nobes:2003nc}
M.~A. Nobes and H.~D. Trottier,
\newblock Nucl. Phys. Proc. Suppl. {\bf 129}, 355 (2004), [hep-lat/0309086].

\bibitem{Trottier:2003bw}
H.~D. Trottier,
\newblock Nucl. Phys. Proc. Suppl. {\bf 129}, 142 (2004), [hep-lat/0310044].

\bibitem{'tHooft:1979uj}
G.~'t~Hooft,
\newblock Nucl. Phys. {\bf B153}, 141 (1979).

\bibitem{Parisi:1984cy}
G.~Parisi,
\newblock Invited talk given at Summer Inst. Progress in Gauge Field Theory,
  Cargese, France, Sep 1-15, 1983.

\bibitem{Adams:2007gh}
D.~H.~Adams and W.~Lee,
\newblock Phys. Rev. D {\bf 77} 045010 (2008), [0709.0781].

\bibitem{Symanzik:1983dc}
K.~Symanzik,
\newblock Nucl. Phys. {\bf B226}, 187 (1983).

\bibitem{vonHippel:2005dh}
G.~M. von Hippel,
\newblock Comput. Phys. Commun. {\bf 174}, 569 (2006), [physics/0506222].

\bibitem{vonHippel:unpub}
G.~M. von Hippel,
\newblock arXiv:0704.0274.

\bibitem{Bernard:2000gd}
C.~W. Bernard {\em et~al.},
\newblock Phys. Rev. {\bf D62}, 034503 (2000), [hep-lat/0002028].

\bibitem{Bernard:2001av}
C.~W. Bernard {\em et~al.},
\newblock Phys. Rev. {\bf D64}, 054506 (2001), [hep-lat/0104002].

\bibitem{Aubin:2004wf}
C.~Aubin {\em et~al.},
\newblock Phys. Rev. {\bf D70}, 094505 (2004), [hep-lat/0402030].

\bibitem{Orginos:1998ue}
MILC, K.~Orginos and D.~Toussaint,
\newblock Phys. Rev. {\bf D59}, 014501 (1999), [hep-lat/9805009].

\bibitem{priv:comm}
D.~Toussaint,
\newblock private communication.

\bibitem{Follana:2007uv}
HPQCD, E.~Follana, C.~T.~H. Davies, G.~P. Lepage and J.~Shigemitsu,
\newblock Phys. Rev. Lett. {\bf 100}, 062002 (2008), [0706.1726].

\end{thebibliography}

\end{document}